\documentclass[notitlepage]{article}

\usepackage{times}
\usepackage{gensymb}
\usepackage{graphicx}
\usepackage{epsfig}
\usepackage{epstopdf}
\usepackage{amsmath}

\usepackage{abstract}

\begin{document}

\title{Global Patterns of Human Synchronization}

\author
{Alfredo J. Morales $^{1,2*}$, Vaibhav Vavilala $^{1}$, Rosa M. Benito $^{2}$, Yaneer Bar-Yam $^{1}$ \\
\normalsize{$^{1}$ New England Complex Systems Institute,}\\
\normalsize{210 Broadway St., Cambridge MA 02139, USA.}\\
\normalsize{$^{2}$ Grupo de Sistemas Complejos, Universidad Polit\'ecnica de Madrid,}\\
\normalsize{ETSI Agr\'onomos, Av. Computlense S/N, 28040, Madrid, Spain.}\\
\normalsize{$^\ast$e-mail: alfredo@necsi.edu}
}

\date{}

\maketitle

\begin{abstract}
Social media are transforming global communication and coordination and provide unprecedented opportunities for studying socio-technical domains. Here we study global dynamical patterns of communication on Twitter across many scales. Underlying the observed patterns is both the diurnal rotation of the earth, day and night, and the synchrony required for contingency of actions between individuals. We find that urban areas show a cyclic contraction and expansion that resembles heartbeats linked to social rather than natural cycles. Different urban areas have characteristic signatures of daily collective activities. We show that the differences detected are consistent with a new emergent global synchrony that couples behavior in distant regions across the world. Although local synchrony is the major force that shapes the collective behavior in cities, a larger-scale synchronization is beginning to occur.
\end{abstract}

The functioning of complex systems, such as human societies or living organisms, arises not only from the individual functionalities of their parts but upon their coordination. 
A central challenge for both sociology and economics is to characterize the way dependencies result in this coordination and the collective activity that comprise our society \cite{bettencourt2013origins}.
Recent studies have shown that these processes can be observed by looking at communication patterns among individuals in social groups \cite{pentland2014social}.
Here we analyze Twitter data to describe the underlying dynamics of social systems. In particular, we study collective activities across geographical scales, from areas smaller than one square kilometer up to the global scale.

The recent explosion of social media is radically changing the way information is shared among people and therefore the properties of our society. These new mechanisms allow people to easily interact with each other and to affordably exchange and propagate pieces of information at multiple scales. As a consequence, people may be able to engage in types of complex tasks previously dominated by organizations structured for a particular purpose \cite{hbr2015}. 
By looking for patterns in the aggregate data, we can retrieve structural and dynamical information about the social system \cite{Lazer2009}. This represents an unprecedented opportunity to study social systems across many scales. 
Traditional surveys of small samples, which are typically limited to a few questions, do not have the scale and frequency to capture such population dynamics.


As highly concentrated social systems, cities are manifestly complex systems with emergent properties \cite{batty2008size}. They are self-organized entities made up of multiple complex agents that engage in larger-scale, complex tasks. Moreover, cities have multiscale structures individually through fractal growth and collectively through size distributions. Their structural patterns have been modeled by scaling laws \cite{bettencourt2013origins}, archetypes of streets layouts \cite{louf2014typology} and land use \cite{decraene2013emergence}. Human generated data has been used to understand the dynamical behavior of inhabitants and their impact on the city functioning. Patterns of human activity and mobility reveal the spatial structure of collective interactions \cite{louail2014mobile} and the dynamical properties of urban functional areas \cite{toole2012inferring}. Many of these studies use mobile phone data which is available only for a small set of cities. 

In this work, we analyze over 500 million geolocated tweets, posted between August 1st, 2013, and April 30th, 2014, to explore patterns of social dynamics in urban areas around the world. We collected these data using the Twitter streaming Application Programming Interface (API) \cite{twitapi2015}, which provides over 90\% of the publicly available geolocated tweets \cite{morstatter2013sample} in real time. Twitter is an online social network whose users share ``micro-blog" posts from smartphones and other personal computers. 
Its population trends younger, wealthier and urban \cite{duggan2013demographics}, which makes it a good probe of the dynamics of young workers in cities. Geolocated tweets 
provide a precise location of the posting individual, and represent around 3\% of the overall tweets \cite{FM4366}. Twitter activity has been analyzed to understand human sentiments \cite{Golder2011}, news sharing networks \cite{Herdagdelen2012} and influence dynamics \cite{Morales2014}, 
as well as global patterns of human mobility \cite{10.1371/journal.pone.0105407}, activity \cite{FM4366} and languages \cite{10.1371/journal.pone.0061981}. 

%

\begin{figure}[ht]
\centering
\includegraphics[width=\linewidth]{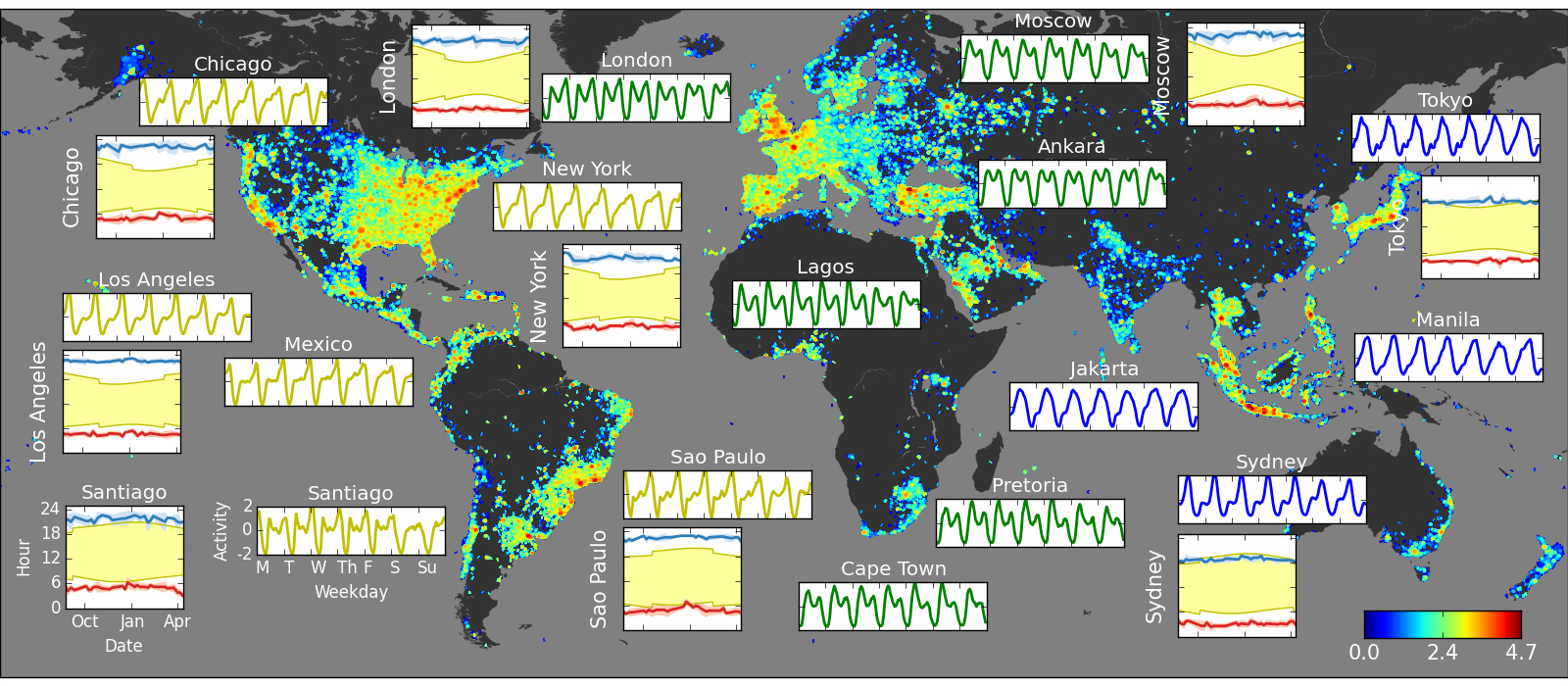}
\caption{Global Twitter Activity. Background map: Twitter activity in each 0.25$^\circ$x0.25$^\circ$ geographic area during an average day (base-10 log scale  at lower right). Rectangular insets: Average week of Twitter activity of selected cities in Universal Time (UTC) after subtracting the mean and normalizing by the standard deviation. Square insets: Low (red) and high (blue) points of Twitter activity of several urban areas compared to daily sunlight periods (yellow) during the nine month observation period (scales on lower left shown for Santiago are the same for all cities). }
\label{fig:ActivityMap}
\end{figure}


In Figure \ref{fig:ActivityMap}, we show the hourly number of tweets during an average week for a few major metropolitan areas (rectangular insets) on top of a map representing the global density of tweets during an average day (52 metropolitan areas across the world are in the Supplement). Overall, cities present similar patterns of collective behavior, cycling between peaks and valleys of activity. Such patterns are also found in phone calls \cite{2008JPhA...41v4015C}, electricity consumption \cite{Pielow2012533} and emails \cite{wang2011information}. Peaks occur during daytime or evening, indicating that people are awake and active, simultaneously tweeting from work, recreation or residential areas, whereas valleys occur during night and sleeping hours, indicating that people are resting and inactive. 

The time series' regular behavior indicates that people synchronize their activities throughout the day. 
This synchrony is not solely due to external factors like light and dark or due to biological factors like circadian rhythms \cite{opac-b1117004}. 
The second set of insets show 
low (red) and high (blue) points of the activity along with the time of sunrise and sunset over the year (yellow shadowed area). The wide range of light and dark times does not cause a comparable shift in activity times 
 (for the most equatorial city $p<0.01$, otherwise $p<0.001$). 
We see that the activity cycles arise instead due to social functions related to economic and social benefits. 
Because our economic system is based upon the contributions of multiple workers, the completion of tasks within a given time frame depends upon the availability of other workers either simultaneously or in the correct sequence \cite{van2004workflow}. 
For workers on a 9-5 schedule, working at the same time every day enables them to meet to conduct business activities together, whether in person or by telephone. Others who work outside of regular business hours are able to provide services like entertainment and shopping opportunities to those who work during the primary shift. 

\begin{figure}[ht]
\centering
\includegraphics[width=\linewidth]{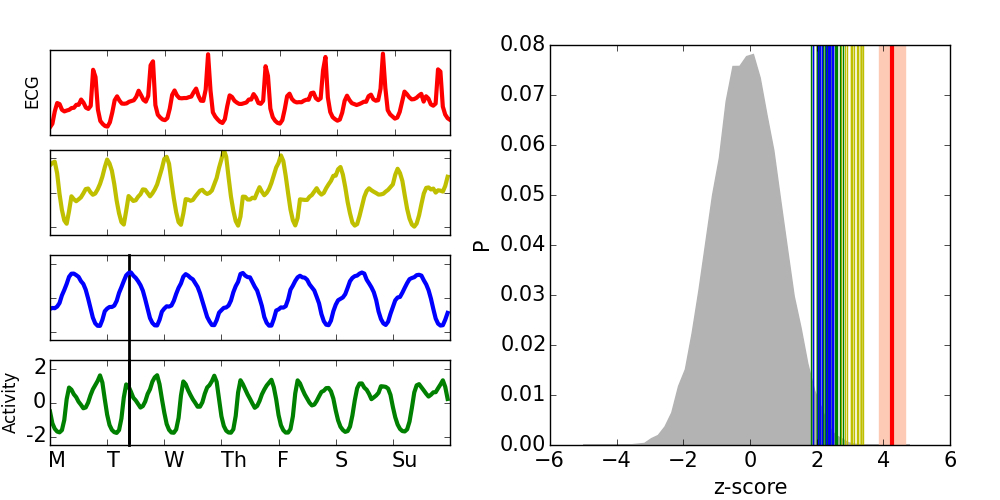}
\caption{ Correlation of the temporal dynamics of cities and heart beats. Left panels: An example of heart beat ECG signal at approximately 80 beats per minute (red) and the average week of activity for three cities: S\~ao Paulo (yellow), Jakarta (blue) and London (green). Vertical black lines show the time of synchronization. Right panel: Correlation of the heartbeat with 50000 random series (gray curve), other heartbeats (red line) and all urban areas colored by group determined by clustering analysis (see Supplement). }\label{fig:HeartBeatCorr}
\end{figure}

Similar patterns arise in the biological activity of living organisms, like heartbeats or respiration, or in their collective activity, like in termite colonies \cite{Jamali20111471} or ecosystems like forests \cite{Grace03111995}. 
Heartbeats in particular have properties that appear in some ways similar to urban dynamics. For comparison, we show in the left panel of Figure \ref{fig:HeartBeatCorr} the electrocardiogram (ECG) of a 43-year-old male \cite{al2000physiobank} together with the Twitter activity from S\~ao Paulo (yellow), Jakarta (blue) and London (green). The similarity between the ECG (red) and the Twitter activity from these cities 
is remarkable, as shown in the right panel of Figure \ref{fig:HeartBeatCorr} (also see Supplement). 
While the high level of correspondence is not essential to our discussion, the reasons for it can be understood. Both regular heart activity and human urban activity have three primary periods. The heart experiences a strong (ventricular) contraction, a secondary (atrial) contraction and a period in which both are relaxing. Human urban collective activity has a primary work shift, a secondary work and recreation shift, and a sleep shift. 

%
%
\begin{figure}[ht]
\centering
\includegraphics[width=\linewidth]{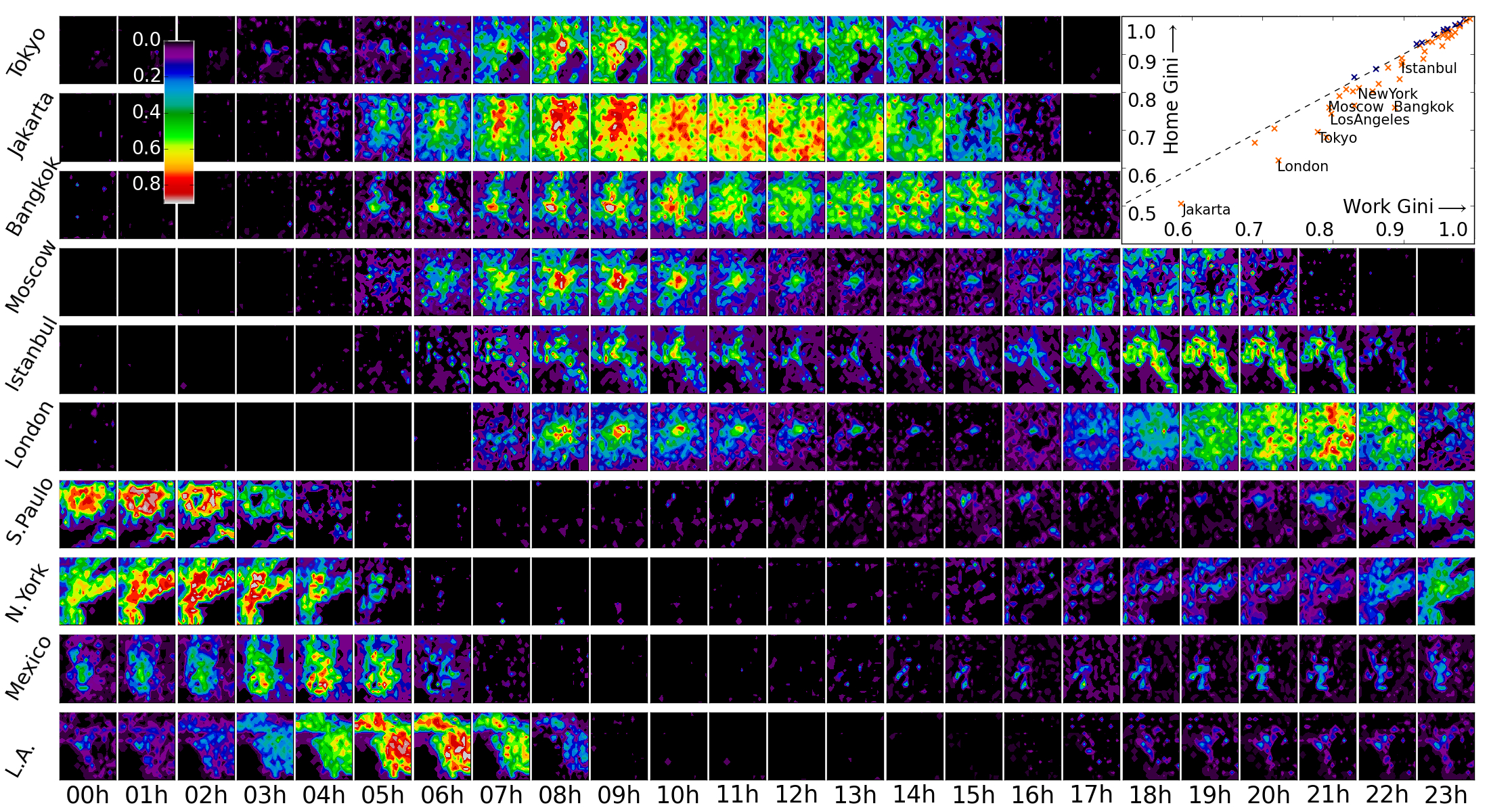}
\caption{Spatio-temporal dynamics of Twitter activity in urban areas. Each row shows activity during an average day according to UTC time for the specified city. Colors indicate the normalized excess of activity from the average value at that location (scale shown in figure). Inset: Geographical heterogeneity (similar to Gini) coefficient of work and home location distributions. }\label{fig:UrbanNebulae}
\end{figure}

People tend to concentrate in a few but dense business or commercial areas during work hours, whereas they disperse to typically sparse residential areas at rest hours. 
We expose this behavior by looking at both spatial variation of Twitter activity and individual mobility patterns (Fig. \ref{fig:UrbanNebulae}). 
We first disaggregated an average day of Twitter activity into a lattice of 20x20 patches in each urban area. In Figure \ref{fig:UrbanNebulae}, we show patches' local activity after subtracting the average, normalizing by the standard deviation and coloring the activity above average. In all cities, there are peaks of intense activity occurring near central areas which expand towards more peripheral areas over time ($p<0.001$). Second, we analyzed the most frequently visited locations of each user. By applying a clustering algorithm \cite{macqueen1967}, two clusters of frequently visited locations were identified (see Supplement). 
In one cluster, most tweets are sent during work hours (after 9am and before 5pm), while in the other cluster most tweets are sent during rest hours (after 8pm and before 2am).
This technique has been 
used to identify work and home locations from mobile phone data \cite{Becker:2013:HMC:2398356.2398375}. In general, work locations are less homogeneously distributed than residential locations ($p < 0.001$ for half of the analyzed cities) which results in a usually higher spatial dispersion (similar to Gini) coefficient (Fig. \ref{fig:UrbanNebulae} inset). 

\begin{figure}[ht]
\centering
\includegraphics[width=\linewidth]{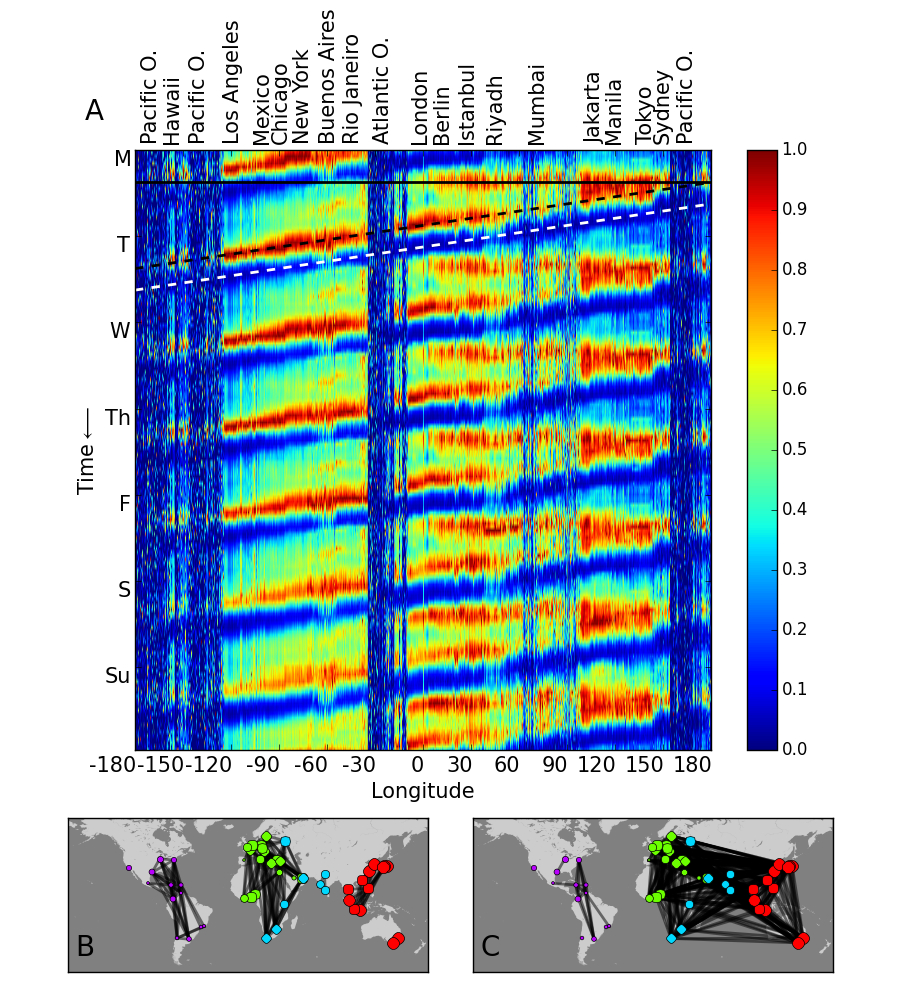}
\caption{A. Temporal dynamics of an average week of Twitter activity by longitude. The vertical axis represents time (increasing from top to bottom) and the horizontal axis represents longitude. Significant cities are indicated at the top. Diagonal dashed lines show peaks of activity (black) and inactivity (white) tracking the time of day. Horizontal line (solid black) indicates synchronous linked activity across Europe, Asia, Africa and Oceania (scale on right). B. Urban correlation network across a time window from 3pm to 3am UTC. C. Like B but across a time window from 1am to 1pm UTC. Colors indicate the results of a clustering algorithm \cite{blondel2008fast}.}\label{fig:WorldDynamics}%
\end{figure}

Despite 
overall similarities, 
cities have distinct signatures in the number and shape of peaks of activity both in time (Fig. \ref{fig:ActivityMap} and \ref{fig:HeartBeatCorr}) and space (Fig. \ref{fig:UrbanNebulae}). One class of cities, including Jakarta (blue series in Fig. \ref{fig:ActivityMap}), has a single large peak of activity during the day. Another class, including S\~ao Paulo (yellow series in Fig. \ref{fig:ActivityMap}), has two small peaks of activity in the morning and a large peak in the evening. Finally, a third class, including London (green series in Fig. \ref{fig:ActivityMap}), has two equally sized peaks of activity, respectively at morning and afternoon. Differences are also manifest spatially (Fig. \ref{fig:UrbanNebulae}). Asian cities (top three rows) gradually increase their activity (colored patches), showing spatial peaks, and rapidly decrease (black patches). European cities (middle three rows) have a strong spatial peak of activity in the morning near the center of the city and other dispersed peaks in the afternoon at peripheral areas. Finally, North and South American cities (bottom three rows) have several smaller peaks of activity around multiple centers and one large peak at the end of the day throughout the city. Interestingly, the morning peaks of the European cities coincide in time with the large afternoon peaks of Asian cities (see columns 8h to 10h), which is an indication of synchrony.
This synchrony is manifested as simultaneous peaks of activity in the time series (vertical black lines in Fig. \ref{fig:HeartBeatCorr} and Supplement A).
 It turns out that some of these differences can be traced to global patterns of behavior.

In Figure \ref{fig:WorldDynamics} A, we show patterns of collective behavior at a global scale after aggregating tweets by longitude. Each longitude has cycles of activity, similar to those of individual cities (Fig. \ref{fig:ActivityMap}, \ref{fig:HeartBeatCorr} and \ref{fig:UrbanNebulae}). Minima (white dashed line) and maxima (black dashed line) shift from east to west due to diurnal synchronization of sleep and wake cycles and the Earth's rotation. The ubiquity of this pattern manifests homogenization in habits and customs among globally differentiated cultures and social contexts. Furthermore, there is a specific synchronization in Figure \ref{fig:UrbanNebulae} that is actually a global phenomenon. Distinct from the other behaviors that track the daily period and therefore are diagonal, between longitudes 0 to 180, a horizontal line (black) shows a simultaneous peak of activity that occurs daily across the European, Asian, African and Oceanian continents. This horizontal peak reflects large scale dependencies across half of the world. 

The synchronization of activity is manifest in a dynamic correlation network between nodes representing cities, whose edges appear when the urban time series are correlated above a given threshold in windows of 12 hours ($r>0.9$). 
In Figure \ref{fig:WorldDynamics} B and C, we show two snapshots of the correlation network at different times. Cities are generally linked within times zones, because it is natural for cities to be synchronized to those in similar longitudes (Fig. \ref{fig:WorldDynamics} B). However, at the time of global synchronization (Fig. \ref{fig:WorldDynamics} C) cities across Eurasia and Africa are strongly coupled, manifesting their synchronous activity (see Supplement for additional evidence from the analysis of Twitter content). 
Such global synchronization can be expected to arise in the context of increasing global interactions consistent with synchronization in many complex systems \cite{boccaletti2002synchronization}.

The patterns we identified manifest how social activities 
lead to dependencies among the actions of individuals, especially their synchronization. Synchronization may arise because of explicit coordination, but more generally occurs due to availability of individuals to perform actions when others trigger responses from them. These couplings might condition people's decisions and diminish individuals' freedom, as they are constrained to the norms and conventions of the social environment. However, as people create and maintain temporally sensitive relationships, the social system increases its complexity at larger scales, enabling collective capabilities that exceed those of an individual.

{\bf Author contributions} 
All authors contributed in making the figures and writing the manuscript.


\begin{thebibliography}{10}

\bibitem{bettencourt2013origins}
Bettencourt, L.~M.
\newblock The origins of scaling in cities.
\newblock {\em Science}{ \bf 340}(6139), 1438--1441 (2013).

\bibitem{pentland2014social}
Pentland, A.
\newblock {\em Social Physics: How Good Ideas Spread-The Lessons from a New
  Science}.
\newblock Penguin, New York, NY,  (2014).

\bibitem{hbr2015}
Libert, B., Wind, Y., and Fenley, M.
\newblock What {A}irbnb, {U}ber, and {A}libaba have in common.
\newblock {\em Harvard Business Review}{ \bf } 11  (2014).
\newblock
  {https://hbr.org/2014/11/what-airbnb-uber-and-alibaba-have-in-common}.

\bibitem{Lazer2009}
Lazer, D., et~al.
\newblock Computational social science.
\newblock {\em Science}{ \bf 323}(5915), 721–--723 (2009).

\bibitem{batty2008size}
Batty, M.
\newblock The size, scale, and shape of cities.
\newblock {\em Science}{ \bf 319}(5864), 769--771 (2008).

\bibitem{louf2014typology}
Louf, R., and Barthelemy, M.
\newblock A typology of street patterns.
\newblock {\em Journal of The Royal Society Interface}{ \bf 11}(101), 20140924
  (2014).
\newblock {doi: 10.1098/rsif.2014.0924}.

\bibitem{decraene2013emergence}
Decraene, J., Monterola, C., Lee, G.~K., Hung, T.~G., and Batty, M.
\newblock The emergence of urban land use patterns driven by dispersion and
  aggregation mechanisms.
\newblock {\em PLoS ONE}{ \bf 8}(12), e80309 (2013).
\newblock {doi: 10.1371/journal.pone.0080309}.

\bibitem{louail2014mobile}
Louail, T., et~al.
\newblock From mobile phone data to the spatial structure of cities.
\newblock {\em Scientific Reports}{ \bf 4}, 5276 (2014).
\newblock {doi: 10.1038/srep05276}.

\bibitem{toole2012inferring}
Toole, J.~L., Ulm, M., Gonz{\'a}lez, M.~C., and Bauer, D.
\newblock Inferring land use from mobile phone activity.
\newblock In {\em Proceedings of the ACM SIGKDD International Workshop on Urban
  Computing},  1--8 (ACM, New York, NY, 2012).

\bibitem{twitapi2015}
{Twitter Streaming Application Programming Interface}.
\newblock {Twitter},  (2015).
\newblock https://dev.twitter.com/streaming/overview [Online; posted 2015].

\bibitem{morstatter2013sample}
Morstatter, F., Pfeffer, J., Liu, H., and Carley, K.~M.
\newblock Is the sample good enough? comparing data from twitter’s streaming
  api with twitter’s firehose.
\newblock In {\em {Proceedings of The 7th International AAAI Conference on
  Weblogs and Social Media }} (The AAAI Press, Palo Alto, CA, 2013).
\newblock
  {http://www.aaai.org/ocs/index.php/ICWSM/ICWSM13/paper/viewFile/6071/6379}.

\bibitem{duggan2013demographics}
Duggan, M., and Brenner, J.
\newblock {\em The demographics of social media users, {P}ew {R}esearch},
  volume~14.
\newblock Pew Research, Washington, DC,  (2013).

\bibitem{FM4366}
Leetaru, K., Wang, S., Cao, G., Padmanabhan, A., and Shook, E.
\newblock Mapping the global {T}witter heartbeat: The geography of {T}witter.
\newblock {\em First Monday}{ \bf 18}(5) (2013).
\newblock {doi: 10.5210/fm.v18i5.4366}.

\bibitem{Golder2011}
Golder, W.~M., and Macy, M.~W.
\newblock Diurnal and seasonal mood vary with work, sleep, and daylength across
  diverse cultures.
\newblock {\em Science}{ \bf 333}(6051), 1878--1881 (2011).

\bibitem{Herdagdelen2012}
Herdagdelen, A., Zuo, W., Gard-Murray, A., and Bar-Yam, Y.
\newblock An exploration of social identity: The geography and politics of
  news-sharing communities in {T}witter.
\newblock {\em Complexity}{ \bf 19}(2), 10--20 (2013).

\bibitem{Morales2014}
Morales, A., Borondo, J., Losada, J.~C., and Benito, R.~M.
\newblock {Efficiency of human activity on information spreading on Twitter}.
\newblock {\em Social Networks}{ \bf 39}, 1--11 (2014).

\bibitem{10.1371/journal.pone.0105407}
Lenormand, M., Tugores, A., Colet, P., and Ramasco, J.~J.
\newblock Tweets on the road.
\newblock {\em PLoS ONE}{ \bf 9}(8), e105407 08  (2014).
\newblock {doi: 10.1371/journal.pone.0105407}.

\bibitem{10.1371/journal.pone.0061981}
Mocanu, D., et~al.
\newblock The {T}witter of {B}abel: {M}apping world languages through
  microblogging platforms.
\newblock {\em PLoS ONE}{ \bf 8}(4), e61981 04  (2013).
\newblock {doi: 10.1371/journal.pone.0061981}.

\bibitem{2008JPhA...41v4015C}
{Candia}, J., et~al.
\newblock {Uncovering individual and collective human dynamics from mobile
  phone records}.
\newblock {\em J. Phys. A}{ \bf 41}(22), 224015 (2008).
\newblock {doi: 10.1088/1751-8113/41/22/224015}.

\bibitem{Pielow2012533}
Pielow, A., Sioshansi, R., and Roberts, M.~C.
\newblock Modeling short-run electricity demand with long-term growth rates and
  consumer price elasticity in commercial and industrial sectors.
\newblock {\em Energy}{ \bf 46}(1), 533 -- 540 (2012).

\bibitem{wang2011information}
Wang, D., et~al.
\newblock Information spreading in context.
\newblock In {\em Proceedings of the 20th {I}nternational {C}onference on World
  Wide Web},  735--744 (ACM, New York, NY, 2011).

\bibitem{opac-b1117004}
Strogatz, S.~H.
\newblock {\em Sync: The Emerging Science of Spontaneous Order}.
\newblock Hyperion {T}heia, New York, NY,  (2003).

\bibitem{van2004workflow}
Van Der~Aalst, W., and Van~Hee, K.~M.
\newblock {\em Workflow Management: Models, Methods, and Systems}.
\newblock MIT {P}ress, Cambridge, MA,  (2004).

\bibitem{Jamali20111471}
Jamali, H., et~al.
\newblock Diurnal and seasonal variations in {C}{H}$_4$ flux from termite
  mounds in tropical savannas of the {N}orthern {T}erritory, {A}ustralia.
\newblock {\em Agricultural and Forest Meteorology}{ \bf 151}(11), 1471 -- 1479
  (2011).

\bibitem{Grace03111995}
Grace, J., et~al.
\newblock Carbon dioxide uptake by an undisturbed tropical rain forest in
  southwest amazonia, 1992 to 1993.
\newblock {\em Science}{ \bf 270}(5237), 778--780 (1995).

\bibitem{al2000physiobank}
Goldberger, A., et~al.
\newblock Physiobank, physiotoolkit, and physionet: components of a new
  research resource for complex physiologic signals.
\newblock {\em Circulation}{ \bf 101}(23), 215--220 Jun  (2000).

\bibitem{macqueen1967}
MacQueen, J.
\newblock Some methods for classification and analysis of multivariate
  observations.
\newblock In {\em Proceedings of the Fifth Berkeley Symposium on Mathematical
  Statistics and Probability, Volume 1: Statistics},  281--297 (University of
  California Press, Berkeley, CA, 1967).

\bibitem{Becker:2013:HMC:2398356.2398375}
Becker, R., et~al.
\newblock Human mobility characterization from cellular network data.
\newblock {\em Commun. ACM}{ \bf 56}(1), 74--82 January  (2013).

\bibitem{blondel2008fast}
Blondel, V.~D., Guillaume, J.-L., Lambiotte, R., and Lefebvre, E.
\newblock Fast unfolding of communities in large networks.
\newblock {\em Journal of Statistical Mechanics: Theory and Experiment}{ \bf
  2008}(10), P10008 (2008).
\newblock {doi: 10.1088/1742-5468/2008/10/P10008}.

\bibitem{boccaletti2002synchronization}
Boccaletti, S., Kurths, J., Osipov, G., Valladares, D., and Zhou, C.
\newblock The synchronization of chaotic systems.
\newblock {\em Physics Reports}{ \bf 366}(1), 1--101 (2002).

\end{thebibliography}
\end{document}